# Методика спільного використання засобів автоматизації лексичного та синтаксичного аналізу в процесі навчання теорії програмування майбутніх учителів інформатики

Сергій Олексійович Семеріков[*], Олександр Павлович Поліщук[≠]
Кафедра фундаментальних і соціально-гуманітарних дисциплін,
ДВНЗ «Криворізький національний університет»,
вул. XXII Партз'їзду, 11, м. Кривий Ріг, 50027, Україна
semerikov@gmail.com[*], apol@cabletv.dp.ua[≠]

**Анотація.** *Цілі дослідження*: розробити методику спільного використання засобів автоматизації лексичного та синтаксичного аналізу lex та yacc у процесі навчання теорії програмування на основі функціональної парадигми.

*Завдання дослідження*: визначити місце і роль синтаксичного аналізу у формуванні професійних інформатичних компетентностей майбутніх учителів інформатики; визначити засоби автоматизації розробки компіляторів у навчанні теорії програмування; розробити основні компоненти методики спільного використання засобів автоматизації лексичного та синтаксичного аналізу в процесі навчання теорії програмування майбутніх учителів інформатики.

*Об'єкт дослідження*: навчання теорії програмування майбутніх учителів інформатики.

*Предмет дослідження*: використання засобів автоматизації лексичного та синтаксичного аналізу в процесі навчання теорії програмування майбутніх учителів інформатики.

Використані *методи дослідження*: аналіз наукових публікацій, самоаналіз досвіду роботи, проектування методики.

*Результати дослідження*. Визначено місце і роль синтаксичного аналізу у формуванні професійних інформатичних компетентностей майбутніх учителів інформатики. Виокремлені засоби автоматизації лексичного (lex) та синтаксичного (yacc) аналізу, інваріантні до використовуваної мови програмування. Показано доцільність використання мов функціонального програмування Scheme та SML для навчання методів розробки компіляторів у курсі теорії програмування. На прикладі діалекту MosML проілюстровано основні компоненти методики спільного використання засобів автоматизації лексичного та синтаксичного аналізу в процесі навчання теорії програмування майбутніх учителів інформатики.

*Основні висновки і рекомендації*:

1) розглянутий приклад розширеного калькулятора може бути

доопрацьований шляхом зміни граматики, зокрема – для уведення умовних та циклічних конструкцій;

2) запропонована схема може бути застосована для реалізації інтерпретатора будь-якої формальної мови з довільним способом типізації – доцільними навчальними прикладами будуть підмножини процедурних мов Basic та C й функціональних Scheme та SML: за умови додавання фази генерації машинного коду це надає можливість продемонструвати повний цикл розробки компілятора мови програмування.

**Ключові слова**: теорія програмування; функціональне програмування; синтаксичний аналіз; SML; Scheme.

### S. O. Semerikov[*], O. P. Polishchuk[ǂ]. Methodic of joint using the tools of automation of lexical and parsing analysis in the process of teaching the programming theory of future informatics teachers

**Abstract**. *Research goals*: to develop a methodic of joint using the tools of automation of lexical (lex) and parsing (yacc) analysis in the process of teaching the programming theory based on a functional paradigm.

*Research objectives*: to determine the place and role of parsing analysis in the formation of professional informatics competences of future informatics teachers; to define tools of compilers development automation in teaching the programming theory; to develop the main components of the methodic of joint using the tools of automation of lexical and parsing analysis in the process of teaching the programming theory of future informatics teachers.

*Object of research*: teaching the programming theory of future informatics teachers.

*Subject of research*: the use of the tools of automation of lexical and parsing analysis in the process of teaching the programming theory of future informatics teachers.

*Research methods* used: analysis of scientific publications, self-analysis of work experience, methodic design.

*Results of the research*. The place and role of parsing analysis in formation of professional informatics competences of future informatics teachers is determined. Separated automation tools for lexical (lex) and syntax (yacc) analysis invariant to the programming language used. The expediency of using functional programming languages Scheme and SML is shown for learning how to develop compilers in the course of programming theory. The example of the MosML dialect illustrates the main components of the methodic of joint using the tools of automation of lexical and parsing analysis in the process of teaching the programming theory of future informatics teachers.

*The main conclusions and recommendations*:

1) the considered example of the expanded calculator can be refined by changing the grammar, in particular – for the introduction of conditional and cyclic constructions;

2) the proposed scheme can be used to implement the interpreter of any formal language with an arbitrary typing method – the appropriate examples of study will be subsets of procedural languages Basic and C and functional languages Scheme and SML: provided the addition of the machine code generation phase, this provides an opportunity to demonstrate the full development cycle for programming language compiler.

**Affiliation:** Department of fundamental and socio-humanitarian disciplines, SIHE «Kryvyi Rih National University», 11, XXII Partz'yizdu str., Kryvyi Rih, 50027, Ukraine.

E-mail: semerikov@gmail.com[*], apol@cabletv.dp.ua[‡].

**Вступ.**

У роботі [2] авторами показано, що розробка програмного забезпечення навчального призначення у підготовці майбутніх учителів інформатики потребує опанування методів аналізу (лексичного, синтаксичного, семантичного) текстів формалізованими мовами та їх інтерпретації.

Класичний курс синтаксичного аналізу включає в себе:

– теорії мов та формальних граматик (класифікація та способи визначення мов, граматики та метамови для їх опису, генерація мов на основі граматик);

– постановку і методи розв'язання задач лексичного аналізу (класифікація лексем та їх виділення із вхідного потоку);

– постановку і методи розв'язання задач синтаксичного аналізу (правильність запису лексем та методи обчислення виразів);

– методи створення і використання програм-генераторів програмних лексичних і синтаксичних аналізаторів;

– програмування інтерпретаторів і компіляторів алгоритмічних мов програмування [2, с. 251].

У роботі [3] подано методику уведення базових понять синтаксичного аналізу, а у роботі [4] – методику використання генератору лексичних аналізаторів lex для реалізації простого калькулятора – інтерпретатору арифметичних виразів, що містять цілі числа, з'єднані операціями додавання та множення; вирази можуть бути згруповані за допомогою дужок.

У роботах [1] та [5] показано, що ефективним засобом опанування початків програмування є функціональний підхід та функціональні мови програмування, такі як слабко типізована мова Scheme та сильно типізована мова SML. Завдяки наявності механізму виведення типів та метапрограмування остання є стає гарним інструментом для навчання курсу «Теорія програмування», до змісту якого й включені питання класичного курсу синтаксичного аналізу. Так, у табл. 1 подано програмну реалізацію алгоритму з [4] мовами Scheme та SML.

*Таблиця 1*

**Порівняльна реалізація простого калькулятора мовами функціонального програмування**

| Scheme | SML |
|---|---|
| ```
#lang racket

(require parser-tools/lex)
(require racket/block)

; tokens
(define INT 1)
(define PLUS 2)
(define MUL 3)
(define OPB 4)
(define CLB 5)
(define QUIT 6)
(define EOL 8)

(define stdin
       (current-input-port))
(define yynumber empty)

(define lexema 0)

(define (E)
 (do [(result (T))]
  [ (not (= lexema PLUS))
    result]
  (set! lexema (yylex stdin))
  (set! result
       (+ result (T)))
 )
)
``` | ```
{
val intOfString = valOf o
Int.fromString;

local open BasicIO Nonstdio
in
datatype tokens =
    PLUS
  | EOF
  | EOL
  | INT
  | OPB
  | CLB
  | QUIT
  | MUL;

val yynumber=ref 0;

val lexema=ref EOF;

fun createLexerStream (is : instream)
= Lexing.createLexer (fn buff => fn n
=> Nonstdio.buff_input is buff 0 n)

fun E(lexbuf)=
 let
  val result=ref 0
 in
  result := T(lexbuf);
  while !lexema=PLUS do (
   lexema := yylex lexbuf;
   result := !result + T(lexbuf)
  );
  !result
 end
and
``` |

| Scheme | SML |
|---|---|
| ```scheme
(define (T)
 (do [(result (F))]
  [ (not (= lexema MUL))
    result]
  (set! lexema (yylex stdin))
  (set! result
        (* result (F)))
 )
)

(define (F)
 (define result 0)

 (if (= lexema OPB)
  (block
   (set! lexema
         (yylex stdin))
   (set! result (E))
   (if (= lexema CLB)
    (set! lexema
          (yylex stdin))
    (display "Error: no )\n")
   )
  )
  (if (= lexema INT) (block
   (set! result yynumber)
   (set! lexema
     (yylex stdin))   )
   (display
    "Error: not a number\n"))
 )
 result
)

(define (calc)
  (display
   "Calc: +, *, (, )\n")
  (set! lexema (yylex stdin))
  (unless (= lexema QUIT)
    (define result (E))
    (display (printf
       "  ~a~n" result))
``` | ```sml
T(lexbuf)=
 let
  val result=ref 0
 in
  result := F(lexbuf);
  while !lexema=MUL do (
   lexema := yylex lexbuf;
   result := !result * F(lexbuf)
  );
  !result
 end
and
F(lexbuf)=
 let
  val result=ref 0
 in
  if !lexema=OPB then
   (
   lexema := yylex lexbuf;
   result := E(lexbuf);
   if !lexema=CLB then
    lexema := yylex lexbuf
   else
    print "Error: no )\n"
   )
  else
   if !lexema=INT then
    (
    result := !yynumber;
    lexema := yylex lexbuf
    )
   else print "Error: not a number\n";
   !result
  end

val _ =
 (while true do
  let val lexbuf = createLexerStream
std_in
   val () = print "Calc: +, *, (, )\n"
  in
   lexema := yylex lexbuf;
   yynumber := E(lexbuf);
   print (Int.toString (!yynumber) ^
"\n")
``` |

| **Scheme** | **SML** |
|---|---|
| ```
    (calc)
   ) )

(calc)

(define yylex (lexer
   [(union "\t" " ")
    (yylex stdin)]
   ["\n" EOL]
  [ (concatenation
    (repetition 0 1 "-")
    (union (concatenation
      (char-range "1" "9")
      (repetition 0 +inf.0
(char-range "0" "9"))) "0") )
   (block (set! yynumber
(string->number lexeme))
INT)]
   ["+" PLUS]
   ["*" MUL]
   ["(" OPB]
   [")" CLB]
   ["quit" QUIT]
   [(eof) QUIT]
)
)
``` | ```
     end)
   handle _ => print "Bye-bye!\n";
end
}

rule yylex = parse
    [` ` `\t`]     { yylex lexbuf }
     (* skip blanks *)
  | [`\n` ]         { EOL }
  | [`-`]?[`0`-`9`]+     {
       (
          yynumber:=
        intOfString (getLexeme lexbuf);
          INT
       )
     }

  | `+`             { PLUS }
  | `*`             { MUL }
  | `(`             { OPB }
  | `)`             { CLB }
  | eof             { EOF }
  | _               { raise Fail
"illegal symbol" }
;
``` |

*Метою дослідження* є розробка методики спільного використання засобів автоматизації лексичного та синтаксичного аналізу lex та yacc у навчанні теорії програмування (на прикладі MosML [6]).

**1. Визначення мови**

Розглянемо приклад розширеного калькулятора із наступною граматичними продукціями $P_{1-17}$:

1) $Defs \rightarrow \varepsilon$
2) $Defs \rightarrow Def\ Defs$
3) $Def \rightarrow \mathbf{id} = Exp$
4) $Exp \rightarrow \mathbf{num}$
5) $Exp \rightarrow \mathbf{float}$
6) $Exp \rightarrow \mathbf{id}$
7) $Exp \rightarrow Exp + Exp$
8) $Exp \rightarrow Exp - Exp$
9) $Exp \rightarrow Exp * Exp$
10) $Exp \rightarrow Exp\ /\ Exp$

11) *Exp* → - *Exp*
12) *Exp* → *Exp* = *Exp*
13) *Exp* → *Exp* and *Exp*
14) *Exp* → *Exp* or *Exp*
15) *Exp* → not *Exp*
16) *Exp* → if *Exp* then *Exp* else *Exp*
17) *Exp* → ( *Exp* )

Виходячи із заданих продукцій, аксіомою граматики є нетермінал *Defs*, використанням якого породжує будь-яку послідовність визначень *Def*. Кожне визначення є прив'язкою до певного ідентифікатора (термінала) `id` за допомогою знаку-термінала = виразу (нетермінала) *Exp*. Вираз може бути терміналами `num`, `float`, `id` або бінарною комбінацією виразів, з'єднаних терміналами +, -, *, /, =, and, or. У двох випадках (продукції 11 та 15) унарні термінали - та not передують виразу, у одному (продукція 17) термінали-дужки ( та ) охоплюють вираз, і ще у одному термінали if, then, else утворюють із нетерміналом *Exp* тернарну операцію.

Таким чином, множину нетерміналів утворюють $V_N$={*Defs*, *Def*, *Exp*}, а термінали складаються із об'єднання таких множин: $V_T$={+, -, *, /, =, (, ), and, or, not, if, then, else} ∪ **num** ∪ **float** ∪ **id**. Останні три множини доцільно визначити через відповідні регулярні вирази. Отже, побудовано граматику G=($V_T$, $V_N$, $P_{1-17}$, *Defs*).

Пріоритети виконання операцій визначимо у такий спосіб:
а) +, -, * та / є лівоасоціативними;
б) * та / мають вищий пріоритет за + и -;
в) унарний мінус має вищий пріоритет за * та /.
г) = є неасоціативним та має пріоритет нижче за + та -;
д) not має більш низький пріоритет, ніж =;
е) and має більш низький пріоритет, ніж not, та є правоасоціативним;
ж) or має більш низький пріоритет, ніж and, та є правоасоціативним.

Крім цих, поширимо поняття пріоритету на else (наскільки це можливо) і зробимо його найменшим із усіх можливих.

**num** визначимо як множину цілочисельних констант, **float** – як множину дійсних констант із плаваючою точкою так само, як й у SML (зокрема, від'ємні числа починатимуться із знаку «~»).

Не входитимуть до граматики, але використовуватимуться такі позасинтаксичні елементи, як однорядкові коментарі (починаються із «\», завершуються кінцем рядка).

Семантика мови полягатиме в тому, що ряд визначень зв'язується із змінними, вирази обчислюються, а результат виводиться.

## 2. Абстракція синтаксису

«Програма» визначеною вище мовою описується абстрактним синтаксисом, тому всі обчислення мають йому відповідати.

Відповідно до продукцій 2 та 3, програма є списком визначень, причому кожне визначення є парою вигляду:

**рядок з іменем змінної** = *вираз*.

Вирази представимо типом даних, який матиме конструктор для кожного типу виразу, крім виразу в дужках. Кожен вираз матимемо, крім підвиразів та атрибутів для імен та чисел, також індикатор положення, який складатиметься з номера рядка та позиції у рядку.

Відповідні оголошення типів згрупуємо у структурі та розмістимо у файлі syntax.sml:

```
structure Syntax =
struct
  type pos = int * int    (* позиція у програмі (рядок, стовпець) *)

  datatype Exp = ICONST of int * pos (* num - цілі сталі *)
     | FCONST of real * pos    (* float - дійсні сталі *)
     | ID of string * pos      (* id - ідентифікатор *)
     | PLUS of Exp * Exp * pos (* + (бінарний плюс) - додавання *)
     | MINUS of Exp * Exp * pos (* - (бінарний мінус) - віднімання *)
     | TIMES of Exp * Exp * pos (* * (бінарна зірочка) - множення *)
     | DIVIDE of Exp * Exp * pos(* / (бінарний слеш) - ділення *)
     | UMINUS of Exp * pos     (* ~ (унарний мінус) - зміна знаку *)
     | EQ of Exp * Exp * pos   (* = - бінарне порівняння *)
     | AND of Exp * Exp * pos  (* and - бінарна логічна кон'юнкція *)
     | OR of Exp * Exp * pos   (* or - бінарна логічна диз'юнкція *)
     | NOT of Exp * pos        (* not - унарне логічне заперечення *)
     | IF of Exp * Exp * Exp * pos (* тернарний умовний вираз *)

  type Def = string * Exp (* визначення - пара виду "назва змінної" = вираз *)
  type Pgm = Def list     (* програма - список визначень *)
end
```

Для компіляції цього файлу скористайтесь командою

mosmlc -c syntax.sml

Результат компіляції буде розміщено у файлах скомпільованого інтерфейсу syntax.ui та об'єктного коду syntax.uo.

## 3. Визначення синтаксису

Отримавши контекстну граматичну специфікацію з додаванням семантичних дій, mosmlyac генерує синтаксичний аналізатор у стилі уасс. Якщо файл parser.grm містить граматичну специфікацію, виклик

mosmlyac parser.grm

створює файл parser.sml, що містить модуль MosML із кодом лексичного аналізатора та файл parser.sig, що містить його інтерфейс.

Згенерований модуль визначає функцію синтаксичного аналізу S для кожного стартового символу *S*, оголошеного в граматиці. Кожна функція синтаксичного аналізу приймає в якості аргументів лексичний аналізатор (функцію перетворення лексичних буферів на лексеми) і лексичний буфер, повертаючи відповідний семантичний атрибут. Функція лексичного аналізатора, як правило, генерується з специфікації лексики за допомого mosmllex. Лексичні буфери є абстрактними типами даних, реалізовані у бібліотечному модулі Lexing. Лексеми є значеннями типу даних token, визначеного в інтерфейсному файлі сігнатури, створеному mosmlyac.

Визначення синтаксичного аналізатору у mosmlyac складається із кількох розділів (деякі з них можуть бути пропущені):

```
%{
  заголовок
%}
оголошення
%%
правила
%%
функції користувача
```

Коментарі в розділах оголошень та правил подаються згідно синтаксису мови C /* та */, тоді як коментарі у розділі заголовка та функцій користувача подаються згідно синтаксису мови ML (* та *).

SML-код, розміщений у заголовку, копіюється у початок файлу parser.sml після визначення типу даних token; він зазвичай містить оголошення open, що вимагаються семантичними діями правил. Будь-який SML-код у розділі функцій користувача копіюється у кінець файлу parser.sml. Обидва ці розділи є необов'язковими.

Оголошення розміщуються по одному на рядок і починаються зі знаку %:

```
%token symbol1 ... symboln
```

Оголосити ці символи як лексеми («токени» – термінальні символи). Ці символи стають конструкторами (без аргументів) у типі даних token.

```
%token < type > symbol1 ... symboln
```

Оголосити ці символи як лексеми з прикріпленим атрибутом даного типу. Ці символи стають конструкторами (з аргументами даного типу) у типі даних token. Тип є частиною довільного типованого виразу MosML, але всі конструктори типу повинні бути повністю кваліфіковані (наприклад, Unitname.typename) для всіх типів, крім стандартних убудованих типів, навіть якщо правильно оголошені open (наприклад,

```
open Unitname
```
) були вказані в розділі заголовку.
```
%start symbol
```

Оголошує даний символ аксіомою (точкою входу, стартовим символом) граматики. Для кожної точки входу визначається функція синтаксичного аналізатору з тим самим ім'ям, визначено у вихідному файлі parser.sml. Для нетерміналів, що не були оголошені як стартові символи, функції синтаксичного аналізатору не генеруються.
```
%type < type > symbol1 ... symboln
```

Визначає тип семантичних атрибутів для заданих символів. Кожен нетермінальний символ, включаючи початкові символи, повинен мати тип його семантичного атрибута, оголошений у такий спосіб. Це гарантує, що згенерований синтаксичний аналізатор є типобезпечним. Тип може бути довільним типом MosML, але всі конструктори типу повинні бути повністю кваліфіковані (наприклад, `Unitname.typename`) для всіх типів, крім стандартних убудованих типів, навіть якщо правильно оголошені `open` (наприклад, `open Unitname`) були вказані в розділі заголовку.
```
%left symbol1 ... symboln
%right symbol1 ... symboln
%noassoc symbol1 ... symboln
```

Визначає пріоритет і асоціативність даних символів. Всі символи, вказані у одному рядку, мають однаковий пріоритет: вищий, ніж символи, оголошені в попередніх визначеннях `%left`, `%right` або `%noassoc`, але нижчий, ніж символи, оголошені в наступних визначеннях `%left`, `%right` або `%noassoc`. Символи оголошуються як лівоасоціативні (`%left`), правоасоціативні (`%right`) або неасоціативні (`%noassoc`). Символи, як правило, є лексемами, але можуть також бути й простими нетерміналами для використання з директивою `%prec` у правилах.

Формат граматичних правил:
```
нетермінал :
     символ ... символ { семантична дія }
   | ...
   | символ ... символ { семантична дія }
;
```

Кожна права частина складається з послідовності символів (можливо, порожньої), за якою слідує семантична дія.

Директива «`%prec символ`» може з'являтися серед символів у правій частині правила для того, щоб вказати, що дане правило має такий самий пріоритет і асоціативність, як і даний символ.

Семантичні дії – це довільні вирази MosML, які обчислюються для отримання семантичних атрибутів, прикріплених до визначеного нетерміналу. Семантичні дії можуть отримати доступ до семантичних

атрибутів символів у правій частині правила за допомогою виразів із `$`: `$1` є атрибутом першого (лівого) символу, `$2` – атрибут другого символу тощо. Порожня семантична дія визначається як `() : unit`.

Для описаної мови визначити лексеми для кожного термінального символу граматики, а також для кінця файлу (EOF). Кожна лексема задається ім'ям та типом. Для більшості лексем тип буде використовуватись лише для визначення позиції, проте для цілих чисел, дійсних чисел та ідентифікаторів тип буде явно містить комбінацію типу MosML та позиції:

```
%token <int*(int*int)> NUM  /* ціле число */
%token <real*(int*int)> FLOAT /* дійсне число */
%token <string*(int*int)> ID /* ідентифікатор */
%token <(int*int)> IF THEN ELSE AND OR NOT EQ
%token <(int*int)> PLUS MINUS TIMES DIVIDE LPAR RPAR EOF
```

Назви визначених лексем відповідають назвам елементів множини термінальних символів, за винятком `EQ` (перевірка на рівність), `PLUS` (додавання), `MINUS` (віднімання), `TIMES` (множення), `DIVIDE` (ділення), `LPAR` (відкриваюча кругла дужка), `RPAR` (закриваюча кругла дужка).

Далі визначимо пріоритет оператору `else` та інших:

```
%nonassoc ELSE       /* блок else тернарного оператору */
%right OR            /* кон'юнкція та диз'юнкція обчислюються, */
%right AND           /* починаючи з правого боку */
%nonassoc NOT        /* для унарних операцій асоціативність */
%nonassoc EQ         /* не визначається */
%left PLUS MINUS     /* арифметичні операції обчислюються */
%left TIMES DIVIDE   /* зліва направо */
%nonassoc UMINUS
```

Порядок рядків має значення – чим нижче розташований рядок, тим вище пріоритет операції. Зверніть увагу на оголошення `UMINUS`: для цього імені немає окремого знаку, але ми збираємось використати його для локальної зміни пріоритету `MINUS`.

Частина операторів оголошена неасоціативними – взагалі-то неважливо, яка асоціативність їм надається, тому що вона немає сенсу для унарних операторів.

Далі оголосимо нетермінали, починаючи з визначення аксіоми граматики та уводячи тип абстрактного синтаксису для всіх інших нетерміналів:

```
%start Defs
%type <Syntax.Pgm> Defs
%type <Syntax.Def> Def
%type <Syntax.Exp> Exp
```

Услід за «%%» визначимо правила граматики. Кожна продукція починається з імені нетерминалу, за яким йде двокрапка, права частина продукції та семантичні дії, згруповані у фігурні дужки. Наступна

продукція для того ж самого нетерміналу не містить його імені та двокрапки – вони замінюються на символ «|». Після завершення всіх пов'язаних із нетерміналом продукцій ставиться крапка з комою:

```
%%
Defs:
    EOF                     { [] }
  | Def Defs                { $1 :: $2 }
;
Def:
    ID EQ Exp               { (#1 $1, $3) }
;
Exp:
    NUM                     { Syntax.ICONST (#1 $1, #2 $1) }
  | FLOAT                   { Syntax.FCONST (#1 $1, #2 $1) }
  | ID                      { Syntax.ID (#1 $1, #2 $1) }
  | Exp PLUS Exp            { Syntax.PLUS ($1, $3, $2) }
  | Exp MINUS Exp           { Syntax.MINUS ($1, $3, $2) }
  | Exp TIMES Exp           { Syntax.TIMES ($1, $3, $2) }
  | Exp DIVIDE Exp          { Syntax.DIVIDE ($1, $3, $2) }
  | MINUS Exp %prec UMINUS  { Syntax.UMINUS ($2, $1) }
  | Exp EQ Exp              { Syntax.EQ ($1, $3, $2) }
  | Exp AND Exp             { Syntax.AND ($1, $3, $2) }
  | Exp OR Exp              { Syntax.OR ($1, $3, $2) }
  | NOT Exp                 { Syntax.NOT ($2, $1) }
  | IF Exp THEN Exp ELSE Exp { Syntax.IF ($2, $4, $6, $1) }
  | LPAR Exp RPAR           { $2 }
;
%%
```

Перше правило гарантує, що список визначень, які складають програму, завершується `EOF`. Також зверніть увагу на продукцію з `UMINUS`: тут виконана локальна зміна вживання лексеми `MINUS` з використанням вказівки `%prec`.

Семантичні дії для продукції, задані між фігурними дужками, є виразами SML, як правило, у визначеній абстракції синтаксису. Символи `$1`, `$2` тощо є спеціальними типами змінних, які містять відповідні семантичні атрибути символів (їх значення) – відповідно першого символу, другого тощо.

Дії для `Defs` будують список визначень: за першою продукцією EOF відповідає порожній список, за другою `Def` визначається на початку списку, що рекурсивно виробляється `Defs`.

Продукція для `Def` визначає ідентифікатор як атрибут, що являє собою пару змінних – ім'я та позицію. Вираз `#1 $1` отримує із цього атрибуту ім'я, яке співставляється із значенням, що є атрибутом нетерміналу `Exp`. Оскільки це третій символ у правій частині продукції (другий – знак рівності), для отримання його атрибуту використовується

$3.

У продукціях для `Exp` є дії, визначені у абстракції синтаксису. зверніть увагу, що перед конструктором типу вказується ім'я модуля `Syntax`. Абстрактний синтаксис для кожного виразу визначає його позицію у програмі – її можна отримати, узявши атрибут `position` для однієї з лексем із правої частини продукції. Більшість лексем мають лише один цей атрибут, тому його можна отримати, використовуючи змінну із знаком долару. Числові сталі (цілі та дійсні) й імена мають власне значення/ім'я у якості атрибуту, тому відповідні частини атрибуту отримуються операторами `#1` та `#2`.

Нарешті, зверніть увагу на дії для виразу в дужках: він повертає лише значення виразу, не беручи дужки до уваги, адже метою їх появи у тексті програми є виключно групування виразів.

Всі визначення синтаксису розміщуються у файлі parser.grm. Синтаксичний аналізатор генерується викликом

mosmlyac -v parser.grm

Цей виклик генерує файли parser.sig, parser.sml та parser.output. Якщо у граматиці є конфлікти, mosmlyac вкаже кількість та тип конфліктів. Для їх деталізації доцільно використати файл parser.output, який містить опис станів та переходів між ними для створеного скінченного автомата, що реалізує висхідний синтаксичний аналізатор у термінах «згортка – перенос».

Для компіляції синтаксичного аналізатору використовуються команди

mosmlc -c parser.sig
mosmlc -c parser.sml

При компіляції parser.sig отримується попередження про те, що використане визначення локального набору виразів є розширенням MosML, яке не відповідає стандарту – його можна проігнорувати. Якщо виникає помилка при компіляції parser.sml, зазвичай вона пов'язана з семантичними діями та визначенням лексем чи нетерміналів.

## 4. Визначення лексики

На основі набору регулярних виразів із прикріпленими семантичними діями mosmllex генерує лексичний аналізатор у стилі lex. Якщо файл lexer.lex містить специфікацію лексичного аналізатора (опис лексики мови), то виконання

mosmllex lexer.lex

генерує файл lexer.sml, що містить MosML-код для лексичного аналізатора. Цей файл визначає одну лексичну функцію на кожну точку входу у визначення лексики. Ці функції мають ті ж самі назви, що й точки

входу. Функції лексичного аналізу приймають у якості параметру лексичний буфер і повертають відповідні семантичні атрибути.

Лексичний буфер – це абстрактний тип даних, який реалізується в бібліотеці `Lexing`. Функції `createLexerString` і `createLexer` з модуля `Lexing` створюють лексичні буфери, які читають відповідно із символьного рядка або будь-якої функції читання.

При використанні разом із синтаксичним аналізатором, створеним mosmlyac, семантичні дії обчислюють значення, що відноситься до типу даних token, визначеного згенерованим синтаксичним аналізатором.

Визначення лексики повинно містити правило, що визначає спеціальний символ `eof`. Крім того, лексичний аналізатор повинен вміти опрацьовувати всі символи, які можуть з'явитися на вході. Це, як правило, досягається шляхом введення символу підстановки _ наприкінці визначення лексики.

Формат визначення лексики такий:
```
{ заголовок }
let abbrev = regexp
...
let abbrev = regexp
rule entrypoint =
  parse regexp { семантична дія }
      | ...
      | regexp { семантична дія }
and entrypoint =
  parse ...
and ...
;
```

Коментарі обмежуються (* та *), як у SML. Абревіатура (`abbrev`) для регулярного виразу може містити лише абревіатури, які передують йому в списку скорочень; зокрема, абревіатури не можуть бути рекурсивними.

Розділ заголовка – це довільний текст SML, який вкладено в фігурні дужки { та }; його можна опустити. Якщо він присутній, прикріплений текст копіюється так, як це було на початку вихідного файлу lexer.sml. Як правило, розділ заголовка містить директиви open, необхідні для дій, і, можливо, деякі допоміжні функції, які використовуються в семантичних діях.

Імена точок входу повинні бути коректними ідентифікаторами SML.

Регулярні вирази `regexp` задаються в стилі lex, але з більш ML-подібним синтаксисом (табл. 2)

*Таблиця 2*

**Подання регулярних виразів у MosML**

| Вираз | Пояснення |
|---|---|
| `` `char` `` | символьна константа |

| Вираз | Пояснення |
|---|---|
| `_` | будь-який символ |
| `eof` | кінець входу лексичного аналізатора |
| `"рядок"` | рядкова константа |
| `[ набір-символів ]` | відповідність будь-якому символу, що належить заданому набору символів. Коректними наборами символів є: одиночні літери `` `с` ``; діапазони символів `` `c_1` `` - `` `c_2` `` (всі символи між $c_1$ і $c_2$, включно); об'єднання двох або більше символьних множин, позначених конкатенацією |
| `[ ^ набір-символів ]` | відповідність будь-якому символу, який не входить до даного набору символів |
| `регулярний-вираз *` | конкатенація нуля або більше рядків, які відповідають регулярному виразу |
| `регулярний-вираз +` | конкатенація одного або декількох рядків, які відповідають регулярному виразу |
| `регулярний-вираз ?` | відповідність порожньому рядку, або рядку, що відповідає регулярному виразу |
| `регулярний-вираз1 \| регулярний-вираз2` | будь-який рядок, який відповідає або регулярний-вираз$_1$, або регулярний-вираз$_2$ |
| `регулярний-вираз1 регулярний-вираз2` | конкатенація двох рядків, першого – регулярний-вираз$_1$, другого – регулярний-вираз$_2$ |
| `абревіатура` | відповідність регулярному виразу в останній let-прив'язці абревіатури. |
| `(регулярний-вираз)` | відповідність регулярному виразу |

Оператори `*` і `+` мають найвищий пріоритет, за яким слідує `?`, потім конкатенація, потім `|` (альтернатива).

Дія (семантична дія) є довільним виразом MosML. Дія виконується в контексті, де ідентифікатор `lexbuf` пов'язаний із поточним лексичним буфером. Деякі типові способи використання `lexbuf` разом із операціями над лексичними буферами (надані бібліотекою `Lexing`) наведено нижче:

`Lexing.getLexeme lexbuf` – повернути рядок, що відповідає поточній лексемі;

`Lexing.getLexemeChar lexbuf n` – повернути n-й символ у поточній лексемі (перший символ має номер 0);

`Lexing.getLexemeStart lexbuf` – повернути абсолютну позицію у вхідному тексті початку поточної лексеми (перший символ, що читається з вхідного тексту, має позицію 0);

`Lexing.getLexemeEnd lexbuf` – повернути абсолютну позицію у

вхідному тексті останнього знаку поточної лексеми (перший символ, що читається з вхідного тексту, має позицію 0);

entrypoint lexbuf – визначити іншу точку входження у тому самому лексичному аналізаторі та рекурсивно викликати його для даної точки входження (корисно для лексичного аналізу вкладених коментарів тощо).

Символьні константи у визначенні лексики для MosML обмежуються символами ` (обернений апостроф), між якими містяться звичайні або спеціальні символи (\\ – обернений слеш, \` – обернений апостроф, \n – новий рядок, \r – повернення каретки, \t – горизонтальна табуляція, \b – видалення попереднього символу, \^c – символ ASCII Control-c, \ddd – символ із десятковим ASCII-кодом ddd).

Рядкові константи задаються послідовністю символів, обмежених символами " (подвійні лапки), або strchar <послідовність символів>.

Вхідний файл для генератора лексичних аналізаторів розмістимо у файлі lexer.lex:

```
{ (* заголовок – вставляється у початок файлу без змін *)
  open Lexing; (* модуль для лексичного аналізу *)

  val currentLine = ref 1 (* номер поточного рядка *)
  val lineStartPos = ref [0] (* список - позиції у рядку *)

  fun getPos lexbuf = getLineCol (getLexemeStart lexbuf)
                                 (!currentLine) (!lineStartPos)
  and getLineCol pos line (p1::ps) =
        if pos>=p1 then (line, pos-p1)
        else getLineCol pos (line-1) ps

  (*визначення виключення у вигляді (повідомлення,(рядок,стовпець))*)
  exception LexicalError of string * (int * int)

  (* функція для опрацювання виключних ситуацій *)
  fun lexerError lexbuf s = raise LexicalError (s, getPos lexbuf)

  (* вибір дії, пов'язаної із ключовим словом *)
  fun keyword (s, pos) =
    case s of
      "if" => Parser.IF pos
    | "then" => Parser.THEN pos
    | "else" => Parser.ELSE pos
    | "and" => Parser.AND pos
    | "or" => Parser.OR pos
    | "not" => Parser.NOT pos
    (* все, що не є ключовим словом, вважатимемо ідентифікатором *)
    | _ => Parser.ID (s, pos);
}
```

```
rule Token = parse
  [' ' '\t' '\r'] { Token lexbuf } (* пропуск пробільних символів *)
  | ['\n' '\012'] {
          currentLine := !currentLine+1;
          lineStartPos := getLexemeStart lexbuf :: !lineStartPos;
          Token lexbuf } (* підрахунок кількості рядків *)
  | "\\" [^ '\n']* { Token lexbuf } (* пропуск коментарів *)
  | ['~']?['0'-'9']+ { case Int.fromString (getLexeme lexbuf) of
                        NONE => lexerError lexbuf "Bad integer"
                      | SOME i => Parser.NUM (i, getPos lexbuf)
                    } (* цілі числа у форматі SML *)
  | ['~']?((['0'-'9']+('.'['0'-'9']* )?|'.'['0'-
'9']+))(['e''E']['+''~']?['0'-'9']+)?
                    { case Real.fromString (getLexeme lexbuf) of
                        NONE => lexerError lexbuf "Bad float"
                      | SOME x => Parser.FLOAT (x, getPos lexbuf)
                    } (* дійсні числа у форматі SML *)
  | (['a'-'z'] | ['A'-'Z']) (['a'-'z'] | ['A'-'Z'] | ['0'-'9'])*  {
                      keyword (getLexeme lexbuf, getPos lexbuf) }
                      (* ключове слово або ідентифікатор *)
  | '+' { Parser.PLUS (getPos lexbuf) } (* прості лексеми *)
  | '-' { Parser.MINUS (getPos lexbuf) }
  | '*' { Parser.TIMES (getPos lexbuf) }
  | '/' { Parser.DIVIDE (getPos lexbuf) }
  | '(' { Parser.LPAR (getPos lexbuf) }
  | ')' { Parser.RPAR (getPos lexbuf) }
  | '=' { Parser.EQ (getPos lexbuf) }
  | eof { Parser.EOF (getPos lexbuf) } (* кінець файлу *)
  | _   { lexerError lexbuf "Illegal symbol in input" }
;
```

MosMLLex надає можливість визначити позицію лексеми у вигляді кількості символів від початку уведення, що не завжди є зручним – більш корисно мати позицію як номер рядка та позицію у цьому рядку, для чого й визначаються додаткові функції getPos та getLineCol, які використовують глобальні змінні currentLine (вказує на поточний номер рядка) та lineStartPos (вказує позиції початку рядків, виміряні як кількість символів від початку тексту). При зчитуванні символу нового рядка currentLine збільшується на 1, а поточна позиція додається у список lineStartPos. Функція getPos повертає пару currentLine та позицію в рядку.

Для повідомлення про лексичні помилки оголошується виключення LexicalError та допоміжна функція lexerError.

Функція keyword використовується для розпізнавання ключових слів: усі алфавітно-цифрові послідовності, що починаються з букви, розпізнаються одним регулярним виразом, тому дана функція допомагає визначити, чи є він ключовим словом або змінною і повертає відповідну

лексему. Зверніть увагу, що ім'я модуля `Parser` присутнє перед іменем кожного термінала – mosmlyac згенерує для нього функцію, параметром якої є лише позиція, у той час як параметром `Parser.ID` є також й ім'я ідентифікатора.

Після секції заголовку йдуть регулярні вирази, що визначають лексеми, та у фігурних дужках – відповідні дії з їх розпізнавання. Вирази розділяються символом «|» і завершуються крапкою з комою. Дії є виразами SML, які повертають значення розпізнаної лексеми. Якщо повертати лексему непотрібно (коли їй не співставлені дії, наприклад, для коментарів та пропусків), необхідно викликати `Token lexbuf` для отримання наступної лексеми. Саме так це робиться у першому рядку розпізнавача для пропуску пробільних символів (пробілу, табуляції та повернення каретки) – зверніть увагу, що одиночні символи знаходяться у обернених, а не звичайних апострофах.

Наступний регулярний вираз стосується символів переведення на новий рядок (символ 12, FormFeed, тут розглядається як різновид такого переведення). Це також пробільні символи, що пропускаються, але перед новим викликом лексичного аналізатору змінні `currentLine` та `lineStartPos` оновлюються.

Далі – опрацювання коментарів: вони складаються із оберненого слешу та продовжуються до наступного переведення рядка. Це можуть бути будь-які символи, крім символу переведення рядка, що відображається у списку символів, який починаються із символу «^» – даний знак інвертує всі подальші елементи списку і може трактуватися як «всі символи, крім наступних».

Далі йдуть справжні лексеми. Перша – це цілочисельні константи, що складаються з непорожньої множини цифр, яким може передувати знак «мінус» (запишемо його тільдою як унарний мінус у SML). Для отримання із рядка числового значення використаємо функцію Int.fromString, як повертає елемент структури `Option` – `NONE`, якщо число завелике або при його перетворенні були помилки, та `SOME`, якщо число було успішно перетворене.

Константи із плаваючою точкою опрацьовуються аналогічно, проте регулярний вираз є більш складним. Зверніть увагу, що регулярні вирази для цілих та дійсних чисел перетинаються. Оскільки шаблон для цілочисельних констант вказано першим, вони розпізнаватимуться коректно. Якщо вирази для цілих та дійсних чисел переставити місцями, жодного цілого числа розпізнано не буде.

Імена ідентифікаторів починаються з літери і продовжуються літерами чи цифрами. Як вже вказувалось, ми використовуємо функцію `keyword` для виокремлення ключових слів.

Далі йде певна кількість однотипних лексем та eof, що вказує на подію кінця файлу. Завершується визначення лексики значенням за замовчанням – шаблону _, який відповідає усім нерозпізнаним (а тому помилковим) символам.

Для генерації лексичного аналізатора виконуються наступні команди:

mosmllex lexer.lex
mosmlc -c lexer.sml

Після успішного виконання mosmllex вкаже кількість станів та дій скінченного автомату для лексичного аналізу. Останнє відповідає кількості визначень у Token регулярних виразів із пов'язаними діями.

### 5. Інтерпретація виразів

Сигнатуру для інтерпретатора визначимо у файлі interpreter.sig:

```
signature Interpreter =
sig
  (* визначення виключення у вигляді (повідомлення, (рядок,
стовпець)) *)
  exception RunError of string * (int * int)
  (* функція інтерпретації пар виразів виду "змінна=значення" *)
  val evalDefs : Syntax.Pgm -> unit
end
```

Згідно наведеного опису, користувачу надається наступний інтерфейс: виключення RunError, яке буде використовуватись для повідомлення про помилки інтерпретації (попередні визначення виключень ParseError та LexicalError використовуються для повідомлень про помилки на етапах синтаксичного та лексичного аналізу), та стартова функція інтерпретації списку виразів.

Для компіляції файлу сигнатури необхідно виконати
mosmlc -c interpreter.sig

Відповідну сигнатурі структуру розмістимо у файлі interpreter.sml:

```
structure Interpreter :> Interpreter =
struct
  (* Для друку повідомлень про помилки використовуйте "raise RunError
(повідомлення,позиція)" *)
  exception RunError of string * (int * int)

  (* позиція - кортеж з номерів рядка та стовпця *)
  type pos = int*int
  (* тип даних, якими буде оперувати інтерпретатор *)
  datatype value = Int of int | Real of real | Bool of bool
  (* наповнимо таблицю пар "змінна-значення" початковими виразами *)
  val vtable=ref [
                  ("pi", Real(3.14159265359)),
                  ("e", Real(2.71828182846)),
```

```
                    ("one", Int(1))
                  ]

   (* функція пошуку у таблиці значення змінної за її ім'ям *)
   fun lookup x [] =
         NONE
     | lookup x ((y,v)::table) =
         if x=y then SOME v else lookup x table

   (* обчислення виразу *)
   fun evalExp exp =  (
     case exp of
        Syntax.ICONST (n,pos) => Int n (* цілі та дійсні сталі
повертаються без змін *)
      | Syntax.FCONST (k,pos) => Real k
      | Syntax.ID  (x,pos) => (* якщо ідентифікатор був створений,
повертаємо його значення *)
         (case lookup x (!vtable) of
           SOME v => v
         | _ => raise RunError ("Unknown variable "^x,pos))
      | Syntax.PLUS (e1,e2,pos) => (* додавання однотипних чисел *)
         (case (evalExp e1, evalExp e2) of
            (Int m,Int n) => Int (m+n)
          | (Real m,Real n) => Real (m+n)
          | _ => raise RunError ("Non-number argument to +",pos))
      | Syntax.MINUS (e1,e2,pos) => (* віднімання однотипних чисел *)
         (case (evalExp e1, evalExp e2) of
            (Int m,Int n) => Int (m-n)
          | (Real m,Real n) => Real (m-n)
          | _ => raise RunError ("Non-number argument to -",pos))
      | Syntax.TIMES (e1,e2,pos) => (* множення однотипних чисел *)
         (case (evalExp e1, evalExp e2) of
            (Int m,Int n) => Int (m*n)
          | (Real m,Real n) => Real (m*n)
          | _ => raise RunError ("Non-number argument to *",pos))
      | Syntax.DIVIDE (e1,e2,pos) => (* ділення - звичайне та націло *)
         (case (evalExp e1, evalExp e2) of
            (Int m,Int n) => Int (m div n)
          | (Real m,Real n) => Real (m/n)
          | _ => raise RunError ("Non-number argument to /",pos))
      | Syntax.UMINUS (e,pos) => (* зміна знаку на протилежний *)
         (case (evalExp e) of
            Int (m) => Int (~m)
          | Real (n) => Real (~n)
          | _ => raise RunError ("Non-number argument to unary -",pos))
      | Syntax.EQ (e1,e2,pos) => (* порівняння однотипних виразів *)
         (case (evalExp e1, evalExp e2) of
            (Int m,Int n) => Bool (m = n)
          | (Real a,Real b) => Bool (a = b)
```

```
        | (Bool m,Bool n) => Bool (m = n)
        | _ => raise RunError ("Non-comparable argument to =",pos))
    | Syntax.AND (e1,e2,pos) => (* кон'юнкція логічних виразів *)
       (case (evalExp e1, evalExp e2) of
          (Bool m,Bool n) => Bool (m andalso n)
        | _ => raise RunError ("Non-logical argument to and",pos))
    | Syntax.OR (e1,e2,pos) => (* диз'юнкція логічних виразів *)
       (case (evalExp e1, evalExp e2) of
          (Bool m,Bool n) => Bool (m orelse n)
        | _ => raise RunError ("Non-logical argument to or",pos))
    | Syntax.NOT (e,pos) =>  (* заперечення логічного виразу *)
       (case (evalExp e) of
          (Bool m) => Bool (not m)
        | _ => raise RunError ("Non-logical argument to not",pos))
    | Syntax.IF (e1,e2,e3,pos) => (*тернарна операція "якщо-то-інакше"*)
       (case (evalExp e1) of
          (Bool m) => if m then (evalExp e2) else (evalExp e3)
        | _ => raise RunError ("Non-logical argument to if",pos))
  )

  (* функція перетворення у рядок для значень типу value *)
  fun toString v=case v of
    Int (a) => (Int.toString a)
   | Real (b) => (Real.toString b)
   | Bool (c) => (Bool.toString c)

  (* співставлення виразу з ідентифікатором *)
  fun evalDef (id,exp) =
  let
     val res=evalExp exp (* обчислення виразу *)
  in
    (
      (* додавання пари "змінна-значення" у таблицю *)
      vtable := (id, res)::(!vtable);
      (* налагоджувальний друк обчисленого значення *)
      print (id^"="^(toString res)^"\n")
    )
  end

  (* інтерпретація всіх виразів у списку *)
  fun evalDefs lst = (app evalDef lst; ())
end
```

Функція `evalDefs` відповідає першим двом продукціям граматики, які визначають вхід інтерпретатору як список виразів виду «змінна=значення». Для виконання такої програми необхідно обчислити кожен вираз із списку, для чого й використовується функція `app`, що застосовує функцію `evalDef` до кожного виразу із списку.

Ідентифікатори та обчислені вирази зберігатимемо у списку `vtable` як елементи пари (змінна, значення). Для визначення можливих типів значень скористаємось граматикою: а) згідно продукції 4, вираз може бути цілим числом; б) згідно продукції 5, вираз може бути дійсним числом; в) згідно продукцій 9-16, вираз може бути логічною змінною. Відповідно було необхідно визначити новий тип даних `value` з трьома конструкторами – `Int`, `Real` та `Bool`. На початку обчислень таблиця повинна бути порожньою, проте у коді вона заповнена трьома ідентифікаторами, що містять значення двох типів.

Тип `pos` традиційно виразимо парою (рядок, стовпець).

Функція `evalDef` обчислює значення виразу `exp`, додає пару (змінна, значення) у початок списку `vtable` та друкує елементи пари. Останнє є необхідним у зв'язку з відсутністю у граматиці продукції, якій співставлена семантична дія виведення. Це зумовлює обов'язковість даної операції співставлення імені змінною з її значенням та надає можливість перевизначати змінні – у списку `vtable` зберігатимуться як поточні, так й попередні значення змінної.

Отримати поточне значення змінної можна за допомогою функції `lookup`, яка повертає елемент NONE структури option, якщо значення не знайдене, та елемент SOME v для першого знайденого значення v.

Оскільки параметром функції `toString` є значення типу `value`, доводиться для кожного конкретного його підтипу викликати відповідну бібліотечну функцію перетворення цілого, дійсного чисел та логічного значення на рядок.

Функція `evalExp` є найдовшою із наведених, адже у ній реалізуються всі останні продукції. Незважаючи на те, що інтерпретатор може оперувати із логічними значенням, увести їх неможливо – у граматиці передбачено лише цілі та дійсні числа як вирази, що обчислюються без змін. Для лексем, що були розпізнані як ідентифікатори, повертається їх значення. Виконання бінарних операцій додавання, віднімання, множення, ділення та унарної операції зміни знаку виконується лише над однотипним числами, у той час як порівнюватися за допомогою знаку рівності можуть всі типи, визначені у `value`. Логічні операції кон'юнкції, диз'юнкції та заперечення виконуються лише над логічними значеннями (найпростіший спосіб їх задати – скористатись операцію порівняння). Тернарна операція «if-then-else» є менш строго типізованою: логічним значенням має бути лише умова, а значення за її істинності та хибності можуть вироблятися будь-якого типу. Наприклад, вхід

```
a=10 \ a - ціле число
b=-1.2 \ b - дійсне число
c=if a=a then a+3 else b*2.0 \ c буде цілим числом,
```

```
d=if not (a=a) then a+3 else b*2.0 \ d - дійсним,
e=if not (a=a) then a+3 else b=b \ а e - логічним
res =
   if 2*2=4 then
     b+d
   else
     a*c
```
дасть наступний результат:
```
a=10
b=~1.2
c=13
d=~2.4
e=true
res=~3.6
```
Для компіляції файлу структури скористайтесь командою:
mosmlc -c interpreter.sml

### 6. Основна програма

Основна програма розміщена у файлі main.sml:
```
structure Main =
struct
  (* зв'язує буфер лексичного аналізатора із вмістом файлу *)
  fun createLexerStream ( is : BasicIO.instream ) =
    Lexing.createLexer (fn buff => fn n =>
                              Nonstdio.buff_input is buff 0 n)

  (* для компіляції файлу: *)
  fun compile filename =
    (* 1) зв'язуємо буфер лексичного аналізатора із вмістом файлу; *)
    let val lexbuf = createLexerStream (BasicIO.open_in filename)
    (* 2) розпочинаємо синтаксичний аналіз із аксіоми граматики *)
        val defs = (Parser.Defs Lexer.Token lexbuf)
    in
      (* 3) обчислюємо список пар "змінна=вираз" *)
      Interpreter.evalDefs defs
    end

  (* функція друку повідомлення про помилку *)
  fun errorMess s = TextIO.output (TextIO.stdErr,s ^ "\n");

  (* ім'я файлу для компіляції - перший аргумент командного рядка *)
  val _ = compile (List.nth(Mosml.argv (),1))
         handle (* опрацювання виключних ситуацій *)
(* коректне завершення синтаксичного аналізатору як виключна ситуація
*)
              Parsing.yyexit ob => errorMess "Parser-exit\n"
(* при синтаксичних помилках конкретизуємо місце, в якому вони сталися
*)
```

```
            | Parsing.ParseError ob =>
                let
                  val Location.Loc (p1,p2) =
                              Location.getCurrentLocation ()
                  val (lin,col) =
    Lexer.getLineCol p2 (!Lexer.currentLine) (!Lexer.lineStartPos)
                in
                  errorMess ("Parse-error at line " ^
                  makestring lin ^ ", column " ^ makestring col)
                end
              (* аналогічно для помилок розпізнавання лексем *)
            | Lexer.LexicalError (mess,(lin,col)) =>
                errorMess ("Lexical error: " ^mess^ " at line " ^
                    makestring lin ^ ", column " ^ makestring col)
              (* та помилок інтерпретації виразу *)
            | Interpreter.RunError(mess,(lin,col)) =>
                errorMess ("Runtime error: " ^mess^ " at line " ^
                    makestring lin ^ ", column " ^ makestring col)
              (* для всіх інших помилок *)
            | SysErr (s,_) => errorMess ("Exception: " ^ s)
end
```

Даний файл опрацьовується командою:

mosmlc -o main main.sml

Результатом її роботи буде виконуваний файл інтерпретатора нашої мови. Для того, щоб процес компіляції та компоновки був успішний, необхідна наявність результатів попередніх дій: інтерфейсних та об'єктних файлів для модулів syntax, lexer, parser та interpreter. Нагадаємо весь процес:

mosmlc -c syntax.sml
mosmlyac -v parser.grm
mosmlc -c parser.sig parser.sml
mosmllex lexer.lex
mosmlc -c lexer.sml
mosmlc -c interpreter.sig
mosmlc -c interpreter.sml
mosmlc -o main main.sml

Основна функція `compile` працює із файлом, ім'я якого передається у якості аргументу командного рядка. У процесі компіляції аналізуються 4 типи помилок – лексичні, синтаксичні, семантичні та системні, кожен із яких представлений відповідним обробником виключної ситуації. Для деталізації повідомлень про помилки використовується функція `errorMess`, а повідомлення містить не лише діагностику, а й збережену на попередніх етапах позицію лексеми. Процес компіляції проходить у три етапи. Для лексичного аналізу використовується лексичний буфер,

пов'язаний із вмістом переданого у командному рядку файлу – його повертає функція `createLexerStream`. Висхідний синтаксичний аналіз починається із аксіоми граматики викликом функції `Defs` з модулю `Parser`, згенерованої mosmlyac. Параметром функції `Defs` є лексема, яку повертає функції `Token` з модулю `Lexer`, згенерована mosmllex. Семантичний аналіз відділено від синтаксичного – інтерпретація побудованого синтаксичного дерева виконується шляхом його редукції у функції `evalDefs` з модулю `Interpreter` – це єдиний модуль, який довелось писати без використання генераторів lex та yacc у дистрибутиві MosML.

**Висновки та рекомендації**

1. Розглянутий приклад може бути розширений та доопрацьований у будь-який спосіб. Так, перша зміна, яку можна запропонувати у граматиці – це зміна продукції 16 на 16$^*$:

16$^*$) *Def* → `if` *Exp* `then` *Def* `else` *Def*

За зміненої продукції з'явиться можливість використовувати умовну конструкцію не в якості виразу в парі «змінна=вираз», а самостійно. При виконанні такої продукції в якості побічного ефекту можуть з'являтися нові змінні. Наприклад:
```
if ((a=b) and (b=c)) then
  d=a*3
else
  d=a+b+c
```
Аналогічно уводиться продукція, що реалізує умовний цикл:

18) *Def* → `while` *Exp* `do` *Defs* `end`

Новий нетермінал `end` є необхідним для того, щоб було можливість між `do` та `end` розмістити послідовність визначень змінних *Defs* – за продукцією 2 вона є обмеженою лише кінцем файлу, тому даний термінал відіграє роль стоп-символу, який сигналізує кінець тіла циклу. Наприклад:
```
\ ітеративне обчислення факторіалу числа 5
i=1
f=1
while not(i=11) do
  f=f*i
  i=i+1
end
```
2. Наведена схема може бути застосована для реалізації інтерпретатора будь-якої формальної мови з довільним способом типізації. Доцільними навчальними прикладами будуть підмножини процедурних мов Basic та C й функціональних Scheme та SML: за умови додавання фази генерації машинного коду це надає можливість продемонструвати повний цикл розробки компілятора мови

програмування.

**Список використаних джерел**

**References (translated and transliterated)**